\pdfoutput=1

\documentclass[runningheads]{llncs}
\usepackage{times}
\usepackage{latexsym}
\usepackage{url}

\usepackage[T1]{fontenc}

\usepackage[utf8]{inputenc}

\usepackage{microtype}

\usepackage{inconsolata}
\usepackage{todonotes}
\usepackage{multirow}
\usepackage[normalem]{ulem}
\useunder{\uline}{\ul}{}
\usepackage{graphicx}
\usepackage{subcaption}
\usepackage{amsmath}
\usepackage[linesnumbered, ruled, vlined]{algorithm2e}
\usepackage{graphicx}
\usepackage{booktabs}
\usepackage{subfig}

\begin{document}
\title{The art of connections: constructing a social network from the correspondence archive of Sybren Valkema}
\titlerunning{The art of connections}

\author{
  {Vera Provatorova}\textsuperscript{1,2}
  \and
  {Carlotta Capurro}\textsuperscript{3} 
  \and
  {Evangelos Kanoulas}\textsuperscript{1}
}
\authorrunning{Provatorova et al.}
\institute{\textsuperscript{1}University of Amsterdam, Science Park 900, 1098 XH Amsterdam\\ \textsuperscript{2}KNAW Humanities Cluster, Oudezijds Achterburgwal 185, 1012 DK Amsterdam\\
  \textsuperscript{3}Utrecht University, Drift 6, 3512 BR Utrecht \\
  \email{\{v.provatorova,e.kanoulas\}@uva.nl \\
  carlotta.capurro@uu.nl}
}

\maketitle
\begin{abstract}
Social network analysis allows researchers to discover insights from connections between people. While the process of building a social network is relatively straightforward for contemporary social media, deriving connections from historical archives remains a challenging task, with every data collection presenting its unique challenges. Our contribution focuses on building and analysing a social network from the correspondence archive of Sybren Valkema (1916-1996), a Dutch glass artist and educator~\cite{dupre_art_2023}. The archive contains both typewritten and handwritten documents in multiple languages, and includes letters from glass artists, art students, art collectors and other agents. We develop an automatic pipeline approach which includes separating handwritten and typed documents, performing text recognition specific to the document modality, extracting names of people from text using named entity recognition, de-duplicating the resulting names to create actor nodes, classifying the actors using entity linking, and, finally, connecting them together and analysing the resulting network. Every part of the pipeline is evaluated against a manual analysis performed by an art historian on a subset of the data collection in order to find out which pitfalls of the automatic approach need to be resolved in future work and, on the contrary, whether using the automatic approach allows to discover any additional insights. The results show strong performance in discovering sender-receiver connections as well as additional meaningful connections in text, with the main challenge being text recognition on scanned pages.
\end{abstract}
\section{Introduction}
Social network analysis (SNA) is analytical framework that allows researchers to uncover insights from connections and interactions between individuals. By examining these connections, SNA can reveal patterns and structures that provide deeper understanding into various social dynamics and relationships. A social network consists of social entities, such as people or organisations, and relations between them, such as information exchange~\cite{jamali2006different}. The entities are commonly referred to as actors, and the connections between them are known as links~\cite{borgatti2024analyzing}.
While the construction of social networks is relatively straightforward when dealing with contemporary social media data, deriving these connections from historical archives presents a unique set of challenges. Each historical dataset is distinct, often involving diverse document types, languages, and varying levels of preservation, which complicates the network construction process.

Our study focuses on the development and analysis of a social network derived from the correspondence archive of Sybren Valkema (1916-1996), a notable Dutch glass artist and educator. This archive, which spans several decades, includes both typewritten and handwritten documents in multiple languages. It contains letters exchanged with other glass artists, art students, art collectors, and various other individuals involved in the art world. The diversity and complexity of this archive provide a rich yet challenging dataset for social network analysis.

To address the challenges presented by this historical archive, we developed an automatic pipeline approach. This pipeline involves several key steps: separating handwritten and typed documents, performing text recognition tailored to the specific modality of the document, extracting names of individuals using named entity recognition (NER), de-duplicating these names to create distinct actor nodes, classifying these actors through entity linking, and finally, connecting the actors to form a comprehensive social network. Each step of this pipeline is designed to handle the nuances of historical documents, ensuring accurate and meaningful extraction of data. The code of the pipeline is available at~\url{https://github.com/vera-pro/Valkema_network}.

Our pipeline is evaluated against a manual analysis conducted by a domain expert on a subset of the archive. This comparative evaluation aims to identify the pitfalls of the automatic approach and suggest future improvements, as well as determine whether the automated method can uncover additional insights that might be overlooked in manual analysis. 

The rest of this paper is structured as follows. In Section~\ref{sec:related-work}, we review related work at the intersection of social network analysis and digital humanities. Section~\ref{sec:methods} describes our pipeline approach for constructing social networks, as well as the network constructed manually by a domain expert. In Section~\ref{sec:evaluation}, we describe the evaluation setup used in our study. Section~\ref{sec:results-discussion} discusses the evaluation results and compares the networks using graph-based metrics. Finally, Section~\ref{sec:conclusion-future} concludes the paper, summarizing the key contributions and outcomes of our study and presenting potential directions for future research.
\begin{table}[]
\begin{tabular}{@{}l|lllll@{}}
\toprule
metric & \# pages, total & \begin{tabular}[c]{@{}l@{}}\# total pages with \\ entities detected\end{tabular} & \begin{tabular}[c]{@{}l@{}}\# typed pages\\ with entities\\ detected\end{tabular} & \begin{tabular}[c]{@{}l@{}}\# handwritten pages\\ with entities\\ detected\end{tabular} & \begin{tabular}[c]{@{}l@{}}\# letters selected for \\ manual network\\  analysis\end{tabular} \\ \midrule
value & 5.8K & 3.3K & 1.9K & 1.4K & 950 \\ \bottomrule
\end{tabular}
\caption{Correspondence dataset statistics}
\label{tab:dataset-stats}
\end{table}
\section{Related work}
\label{sec:related-work}
Extracting social networks from unstructured text is a complex task at the intersection of knowledge mining, natural language processing, and, when applied to cultural heritage data, digital humanities. Network analysis in digital humanities offers new ways to explore and understand historical and cultural data. Researchers have used it to visualize relationships in historical documents, providing a nuanced interpretation of social and cultural dynamics that traditional methods might overlook. Painter et al.~\cite{painter2019network} discuss the principles, problems, and extensions of network analysis in digital humanities, emphasizing its growing importance and complexity in handling historical data. Besnier~\cite{besnier2020history} demonstrates how social network analysis can trace the evolution of historical accounts into myths, providing valuable insights into cultural storytelling dynamics. Geerlings~\cite{geerlings2015visual} explores the correspondence archives of Rosey E. Pool, using visualization techniques combined with biographical data to provide insights into the social dynamics of historical figures without employing quantitative network metrics. Additionally, Capurro and Severo~\cite{capurro2023mapping} examine the dynamics of digital heritage networks, offering insights into the political and cultural discussions shaping digital heritage. Their study on mapping Europeana's web-based network informs our methodological framework by illustrating the importance of comprehensive digital network analysis in understanding social dynamics within archival content.

 Two contributions most similar to ours are by Dekker et al.~\cite{dekker2019evaluating} and Fields et al~\cite{fields2023using}. In particular, Dekker et al. focus on using named entity recognition (NER) to build a social network from novels, highlighting the importance of developing culturally aware AI systems. This work evaluates several NER tools to extract and analyze character networks within literary texts, emphasizing the need for sensitivity to narrative context and cultural nuances that influence entity recognition and relationship interpretation \cite{dekker2019evaluating}. Our work extends these methods to a broader archival dataset that includes multilingual and multimodal documents from Sybren Valkema's correspondence. 
Similarly, Fields et al. utilize NER and network analysis to differentiate personal relationships from the broader social milieu in historical personal diaries. Their methodology, which considers the socio-historical context of the 19th-century Ottoman-Iraqi region, helps distinguish between various types of social connections, such as family ties and professional networks \cite{fields2023using}. Inspired by their approach, our research seeks to uncover hidden networks in Valkema’s extensive correspondence, exploring both personal and professional dimensions. 
\begin{figure}
    \centering
\includegraphics[width=1\linewidth]{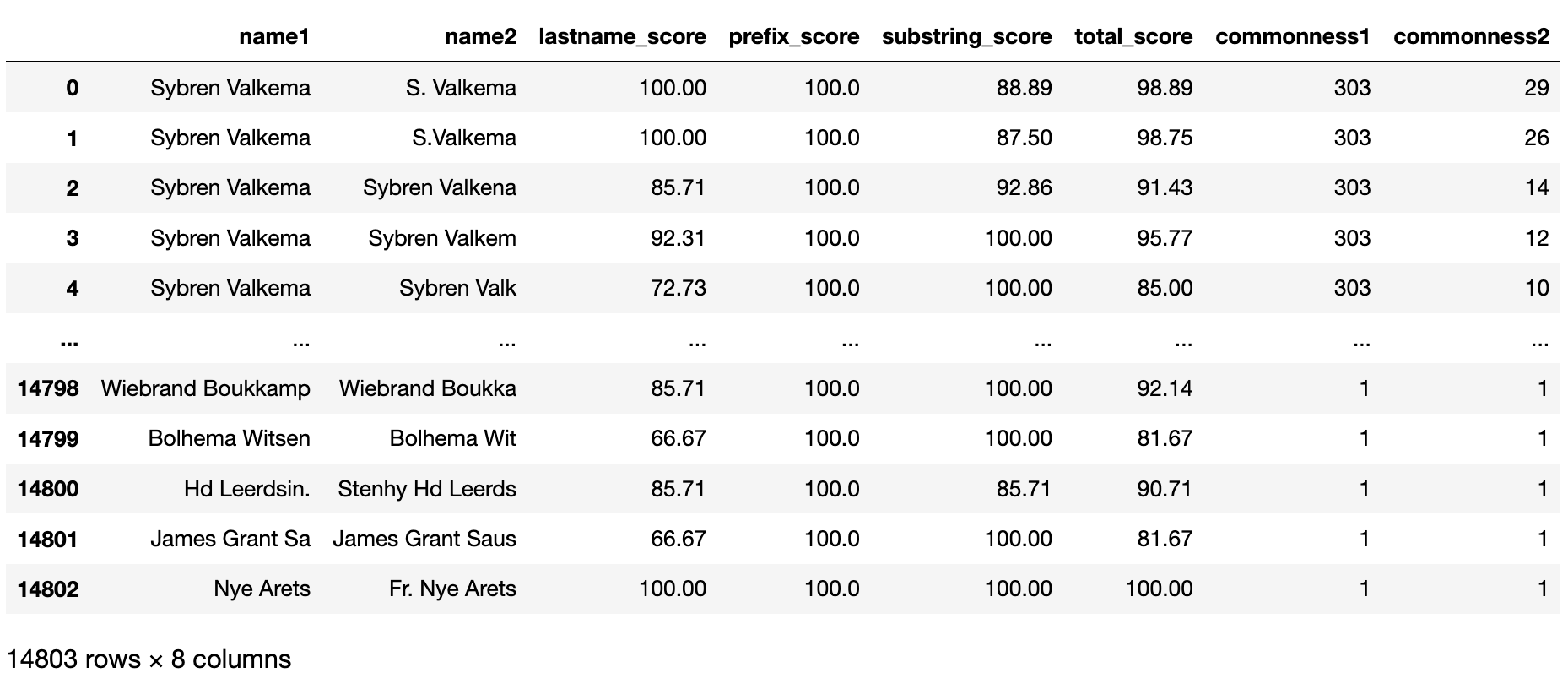}
    \caption{Example of pairwise string similarity scores calculated for record linkage (Step 3 of the pipeline).}
    \label{fig:record_linkage}
\end{figure}
\section{Dataset}
\label{sec:dataset}
The dataset we are focusing on is the digitised archive of Sybren Valkema (1916-1996), a Dutch glass artist and educator who was one of the pioneers of the Studio Glass movement in Europe. The archive contains scanned pages of correspondence, books, drawings, and other documents belonging to the artist. For the purpose of this paper, we are focusing on the correspondence to analyse the connections between Valkema and other people involved in Studio Glass, such as artists, students, museums and art galleries, editors, and art collectors. To separate correspondence from other documents in the archive, we use the metadata: a document is selected if its metadata contains the word "letter" or "correspondence". The dataset statistics are shown in Table~\ref{tab:dataset-stats}.

The dataset presents various challenges. Firstly, the data is highly heterogeneous: while the metadata of every page allows us to find out whether the page is correspondence, false positives are possible, with blank pages, drawings, and booklets sometimes labelled as correspondence. Secondly, the scanned pages are stored as images, with some of the documents being typed and some handwritten. Lastly, the data is multilingual, with most documents written in Dutch, English, and German, as well as a considerable number of pages in French, Swedish, and other languages (e.g., Japanese and Czech). Since the metadata does not contain any information on the language nor modality (typed/handwritten/drawing) of a page, extra effort is needed to prepare the data for analysis.
\begin{figure}
    \centering
    \begin{subfigure}[t]{0.47\textwidth}
        \centering
        \includegraphics[width=\linewidth]{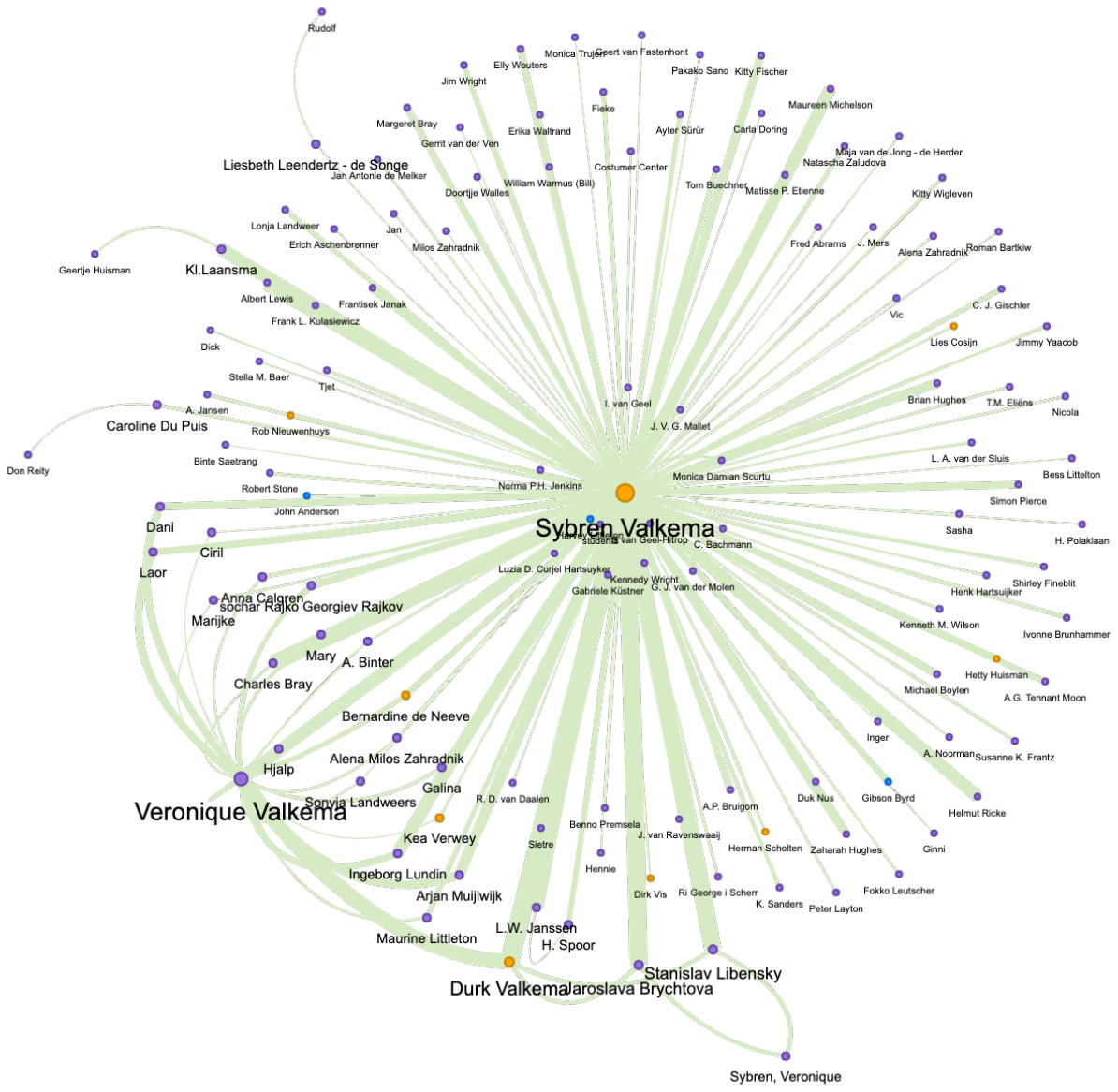}
        \caption{Sender-receiver network $G_{manual}$, constructed manually by domain expert}
        \label{fig:g_manual}
    \end{subfigure}
    \hfill
    \begin{subfigure}[t]{0.49\textwidth}
        \centering
        \includegraphics[width=\linewidth]{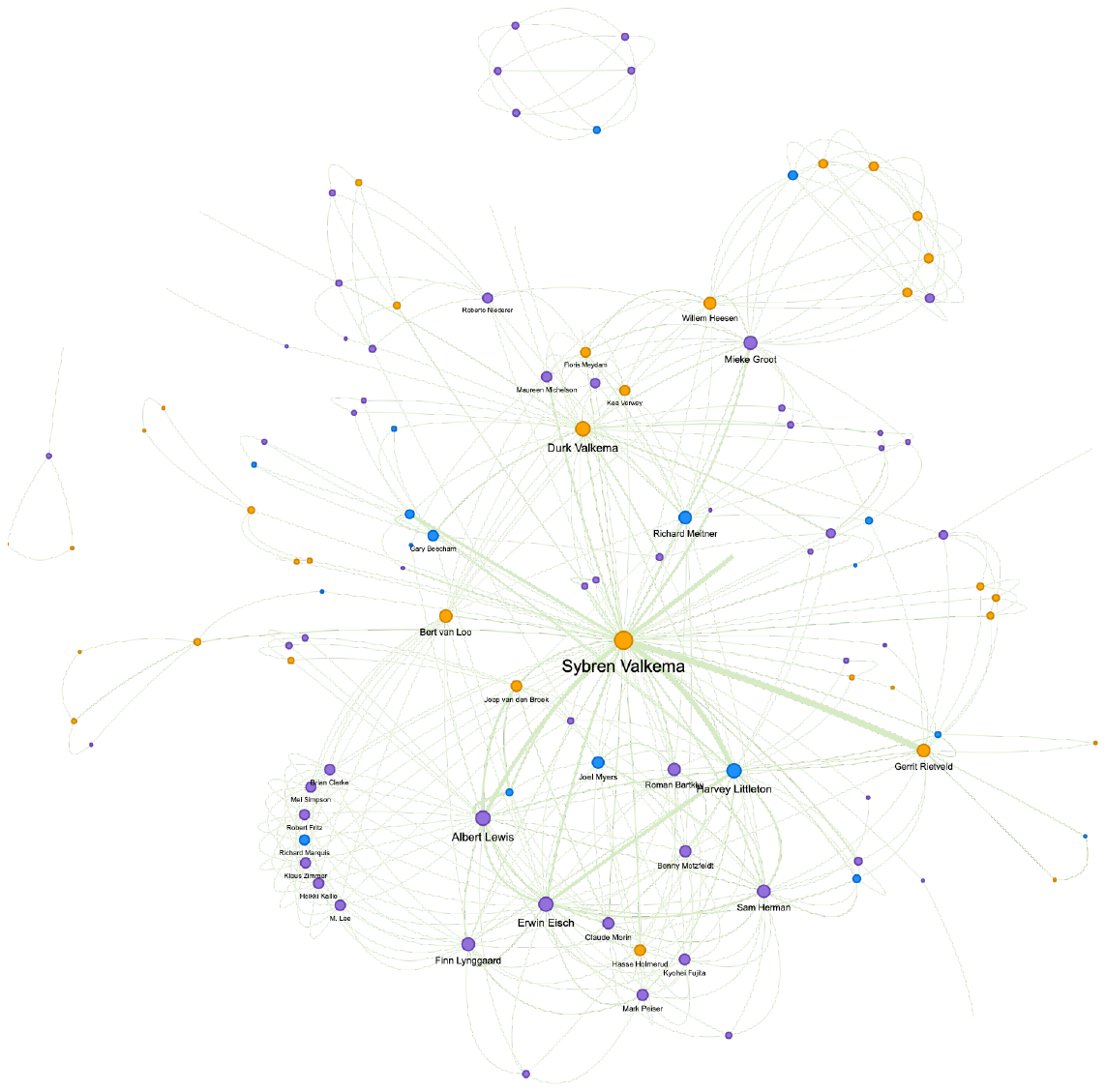}
        \caption{Co-occurrence network $G_{small}$,constructed automatically from the same data sample as $G_{manual}$.}
        \label{fig:g_small}
    \end{subfigure}
    \vspace{0.5cm} 
    \begin{subfigure}[t]{1\textwidth}
        \centering
        \includegraphics[width=\linewidth]{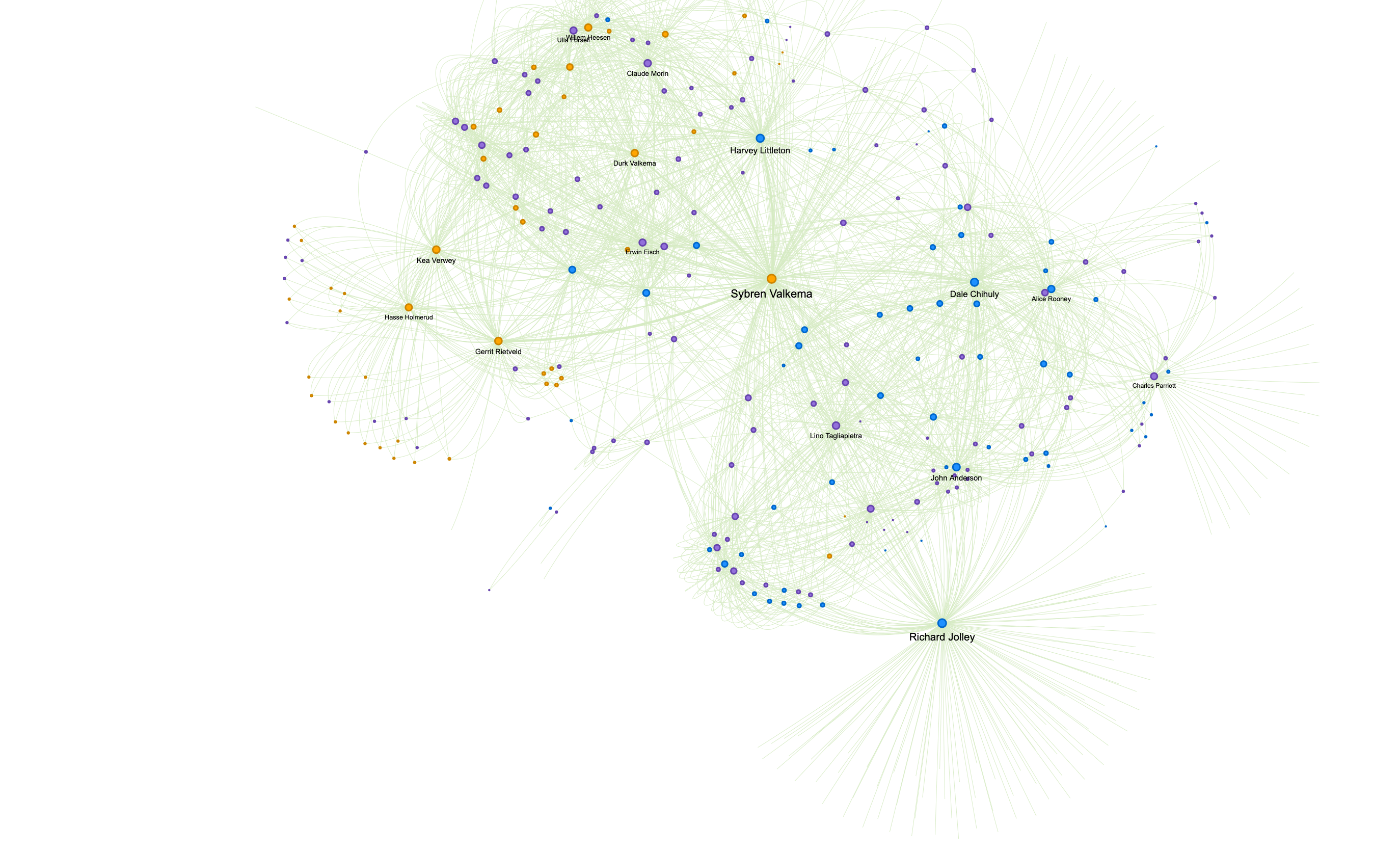}
        \caption{Network $G_{auto}$, constructed automatically on the whole set of correspondence.}
        \label{fig:g_auto}
    \end{subfigure}
    \caption{Visualisation of the three networks constructed from the correspondence archive. Node colours reflect countries of residence retrieved for the actors: orange is the Netherlands, blue is the USA and all remaining nodes are purple.}
    \label{fig:three_subfigures}
\end{figure}
\section{Methods}
\label{sec:methods}
In order to construct a social network from scanned pages of the correspondence archive, we develop a pipeline approach that consists of 5 steps: text recognition, named entity recognition, record linkage, entity linking, and, finally, network construction. To evaluate the pipeline, we compare it with a sender-receiver network constructed manually by a domain expert on a subset of data. This section describes every step of the automatic pipeline and the manual network construction.
\subsection{Text recognition}
Since typed and handwritten pages need to be processed differently, with optical character recognition (OCR) and handwritten text recognition (HTR), respectively, the first step of our pipeline is separating them from each other. We combine the separation process with running OCR on typed pages using the following heuristic: if the Tesseract OCR engine\footnote{\url{https://github.com/tesseract-ocr/tesseract}} fails to detect Latin script on a page, this means that the page is either handwritten or a drawing. Otherwise, when Latin script is detected, the page is marked as typed and the OCR results are saved. Next, for handwritten pages, we perform HTR using Transkribus\footnote{\url{https://www.transkribus.org/}} with Text Titan I\footnote{\url{https://readcoop.eu/introducing-transkribus-super-models-get-access-to-the-text-titan-i/}}, a pre-trained large multilingual handwritten text recognition model. 
\subsection{Named entity recognition}
In order to construct a network from the documents, we start with automatically detecting people in text using named entity recognition (NER). We use three NER models: two of them were fine-tuned on historical Dutch texts and are based on BERTje~\cite{de2019bertje} and mBERT~\cite{devlin2019bert} respectively, while the third model, WikiNEuRal~\cite{tedeschi-etal-2021-wikineural-combined}, is multilingual and has been trained on contemporary texts. The choice of models aims to achieve a balance between historical and contemporary texts as well as Dutch and multilingual texts. While the historical models were fine-tuned on the data older than the 20th century, which is the time span of the Valkema archive, prior work shows that they still perform competitively when recognising entities on the early 21st century data~\cite{provatorova2024too}.

The results are aggregated with the goal of maximising recall: an entity is selected if at least one model predicted it. This way, we aim to extract as many entities as possible from the documents, with eventual false positives being filtered out on the next steps of the pipeline.
Applying entity recognition resulted in detecting 20442 unique entities.
\subsection{Record linkage}
In order to make the network analysis precise, we need to make sure that every node refers to one person. This requires further post-processing of the NER results, since one person can be referred to as multiple entity mentions: for example, "S.Valkema" and "Dhr S. Valkema" both refer to Sybren Valkema. Moreover, OCR and HTR noise in the documents contribute to adding extra spelling variations to the names of actors. We perform record linkage to merge the entity mentions that refer to the same person. To do so, we start with calculating pairwise similarity scores for all entity mentions using string matching and saving the results in a lookup table, as shown on Figure~\ref{fig:record_linkage}. Next, for every entity mention we find its closest match by string similarity. If an entity mention $m_1$ and its best match $m_2$ have a similarity score above the threshold value, they are merged into one entity and saved in an alias dictionary. When choosing which of the two name variations is selected as the main form for an entity, we use commonness scores and length of the names. The assumption is that the canonical form of an entity appears more frequently in the corpus than its variations, and when the number of occurrences is equal, a longer form tends to be more informative. 

To measure similarity between names, we first use a greedy algorithm to split every name into prefix and last name, taking into account the infixes ('van', 'de', etc.) often present in Dutch family names. Then, we calculate 4 string similarity measures: $lastname\_score$, $prefix\_score$, $substring\_score$ and $total\_score$. All of these measures are based on Levenstein distance and calculated with the rapidfuzz\footnote{\url{https://github.com/rapidfuzz/RapidFuzz}} Python library: $lastname\_score$ and $prefix\_score$ are $fuzzy\_ratio$ applied to last names and prefixes respectively, $substring\_score$ is $partial\_ratio$ applied to names without splitting, and $total\_score$ is a weighted average of the three scores. By using these scores, we attempt to take into account various cases of record linkage: for example, using $prefix\_score$ alongside $lastname\_score$ ensures that "D. Valkema" is closer to "Durk Valkema" than to "Valkema". If $lastname\_score$ is high but other scores are not, the names are not matched, which reduces ambiguity: for instance, "Valkema" can refer to Sybren Valkema, his wife Veronique, or his son Durk, also a glass artist. 

The process of record linkage has to deal with extra challenges specific to the dataset. Sybren Valkema was also known as Iep Valkema, which is impossible to find using string matching alone. This connection is added manually at the beginning of the process, ensuring that other forms such as "I. Valkema" later get linked to "Sybren Valkema" automatically.

In total, applying this step of the pipeline has resulted in reducing the number of unique entities from 20442 to 15364.
\subsection{Entity linking}
After initial record linkage, the next step is performed: linking the resulting entities to Wikidata. This step is beneficial for the network analysis since it allows to classify the nodes by their Wikidata properties, making the network more useful for art history research. For instance, retrieving the country of residence for the actors allows to analyse the exchange of information between American and European glass artists. Moreover, entity linking helps to address privacy concerns: some of the actors used to be Valkema's students but then decided to switch careers, and they might prefer to be left out of the network analysis, unlike famous artists and art collectors whose names can be found on Wikidata. 

The process of entity linking consists of two steps: candidate generation and disambiguation. For the first step, we use the Wikidata SPARQL endpoint\footnote{\url{https://query.wikidata.org/sparql}} to retrieve all possible entity identifiers matched by label with every person name in our dataset. To reduce the number of API calls, we start with the canonical form of every name (identified on the previous step of the pipeline). If no matching entities are found for the canonical form of a name, we repeat the query using alternative spellings connected with this name during record linkage, until a candidate is found or all alternative spellings are tried. If no candidates are found, the entity is discarded and not used for constructing the social network.

For the disambiguation step, we first manually disambiguate Sybren Valkema as \url{https://www.wikidata.org/wiki/Q2618110}. Next, we use this information to disambiguate all other entities. For every entity, we retrieve its "description" and "occupation" properties: in case of Sybren Valkema, the values are "Dutch glass artist (1916-1996)" and "glass artist, ceramicist, textile artist" respectively. After this, we use a transformer-based sentence similarity model~\cite{reimers-2019-sentence-bert} to calculate vector embeddings of description and occupation for every entity. Then, for every candidate entity we calculate cosine similarity between its description embedding and the description embedding of Sybren Valkema. Same is done for occupation embeddings, after which the similarity scores are averaged and the candidate with the highest score is selected for the canonical form of every name.
Applying this step of the pipeline has resulted in reducing the number of unique entities from 15364 to 5621.
\subsection{Automatic network construction}
Constructing a social network from entity mentions in correspondence can be approached in two ways: either by adding an edge between any two individuals mentioned together in a single document or by connecting only senders and receivers. For our automatic method, we chose the first approach, as it allows to maximise recall and potentially reveals more insights. In this method, two entities are connected if they appear within the same document, regardless of their position, as this allows us to capture a broader range of interactions. This approach differs from more widespread methods that capture entity co-occurrence within sentences or paragraphs, which is not suitable for our dataset. For example, in our case, the receiver may be mentioned at the beginning of a letter and the sender at the end, and for lengthy documents this means that the connection will be overlooked when detecting co-occurrence on the paragraph level. Additionally, our dataset includes not only letters but also other kinds of documents, such as newspaper articles, brochures, and booklets enclosed to the letters. Our hypothesis is that two artists being mentioned together in the same article or the same letter often implies a professional relationship or collaboration, making these connections meaningful. We test this assumption in manual evaluation conducted by a domain expert by closely reading the documents and certifying the nature of the relationship between the two entities.

We measure the strength of connection between two actors by the frequency of their co-occurrence across the dataset. Thus, the weight of an edge between two nodes $p_1$ and $p_2$ is the number of documents that contain  $p_1$ and $p_2$ together. 
Figure~\ref{fig:g_auto} presents a visualisation of the resulting graph, illustrating the complex web of connections among individuals in Sybren Valkema's correspondence archive. This visualisation includes both prominent relationships and less obvious connections discovered by our analysis. To reduce noise and focus on the most significant relationships, we prune edges with weights below the threshold of 10. 
\subsection{Manual network construction}
Our contribution includes a sender-receiver network constructed manually by a domain expert on a sample of documents from the correspondence archive. This is crucial for identifying both the pitfalls and the benefits of the automatic approach. The sender-receiver network was built using only the letters in the collection. Therefore, the analysis excluded all the other kinds of documents that are labelled as correspondence by metadata but were sent or received together with a letter. The network was built using 950 documents and certified the direct connection existing between the sender and the receiver. Figure~\ref{fig:g_manual} shows a visualisation of the manually constructed network. 
\subsection{Network analysis}
After creating the two networks and evaluating the pipeline, we turn to network analysis. First, we construct an additional network $G_{small}$ (Figure~\ref{fig:g_small}) by applying the same automatic pipeline used to construct the network on all data ($G_{auto}$) to the same data subset that was used for the manually curated sender-receiver network ($G_{manual}$). This is done to minimise the impact of the significant size differences between $G_{auto}$ and $G_{manual}$. Next, we calculate the standard graph-based metrics to compare the three networks (Table~\ref{tab:graph-stats}), and construct centality profiles of the most prominent nodes (Tables~\ref{tab:centrality-auto},~\ref{tab:centrality-manual},~\ref{tab:centrality-small}). The network metrics are calculated using the networkx\footnote{\url{https://networkx.org/}} Python library, and the networks are visualised using pyvis\footnote{\url{https://pyvis.readthedocs.io/en/latest/}}.
\begin{figure}
    \centering
\includegraphics[width=1\linewidth]{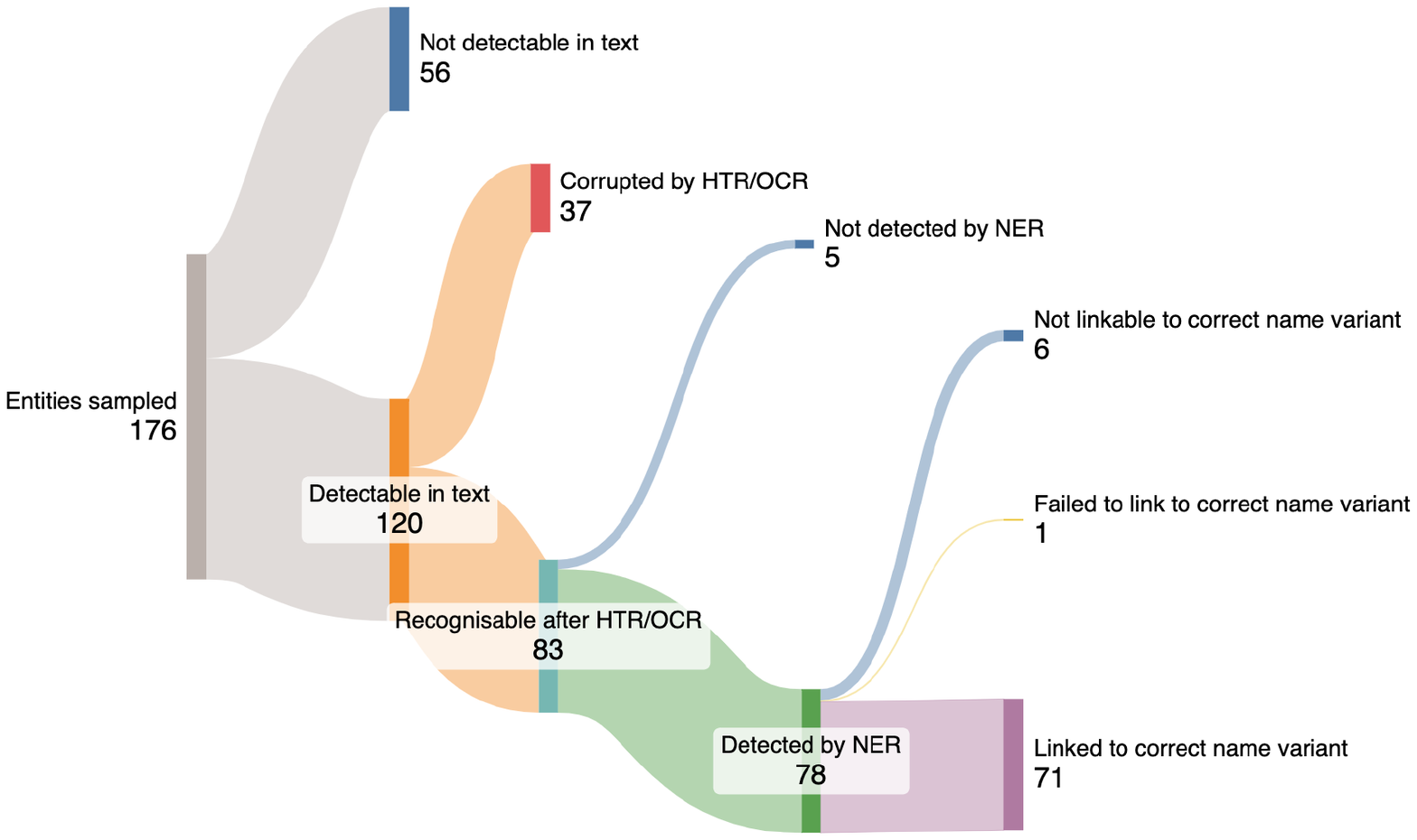}
    \caption{Results of manually evaluating the pipeline on a sample of data.}
    \label{fig:pipeline_sankey}
\end{figure}
\begin{table}[ht!]
\scriptsize
\begin{tabular}{@{}lll@{}}
\toprule
\textbf{error} & \textbf{explanation} & \textbf{example} \\ \midrule
\multicolumn{1}{l|}{Not detectable in text} & \multicolumn{1}{l|}{\begin{tabular}[c]{@{}l@{}}The person’s name is not explicitly\\  mentioned in the text and \\ was inferred by the domain expert \\ from the context and/or the metadata\end{tabular}} & \begin{tabular}[c]{@{}l@{}}A letter from Kea Verwey \\ contains only her signature\\ instead of full name\end{tabular} \\ \midrule
\multicolumn{1}{l|}{Corrupted by HTR/OCR} & \multicolumn{1}{l|}{\begin{tabular}[c]{@{}l@{}}The name is present in the OCR/HTR output\\  but looks unrecognisable \\ to the human annotator due to noise\end{tabular}} & \begin{tabular}[c]{@{}l@{}}In a handwritten letter, the sender's\\ name Ingeborg was recognised as\\ "Sugeboy"\end{tabular} \\ \midrule
\multicolumn{1}{l|}{Not detected by NER} & \multicolumn{1}{l|}{\begin{tabular}[c]{@{}l@{}}The name is present in the OCR/HTR output \\ and looks clear to the human annotator, \\ but is not recognised as a named entity \\ by the pipeline\end{tabular}} & \begin{tabular}[c]{@{}l@{}}In a letter Sybren Valkema is \\ mentioned as "Valkema Sybren" \\ and not recognised as an entity\end{tabular} \\ \midrule
\multicolumn{1}{l|}{Not linkable to correct name variant} & \multicolumn{1}{l|}{\begin{tabular}[c]{@{}l@{}}a) The "main" name variant indicated\\ by the domain expert is not present \\ in the corpus and therefore not possible \\ to connect with the name in text\\ \\ b) The name variant present in text is ambiguous\end{tabular}} & \begin{tabular}[c]{@{}l@{}}a) ‘K. Fischer’ is present in the corpus \\ but ‘Kitty Fischer' is not\\ \\ b) “Heer Valkema” (mister Valkema) \\ could refer to Sybren Valkema \\ or Durk Valkema\end{tabular} \\ \midrule
\multicolumn{1}{l|}{Failed to link to correct name variant} & \multicolumn{1}{l|}{\begin{tabular}[c]{@{}l@{}}The "main" name variant indicated \\ by the domain expert is present in the corpus \\ but not recognised as close enough to the name \\ by the pipeline\end{tabular}} & \begin{tabular}[c]{@{}l@{}}"Robert H. Barber" was not linked\\ to "Robert Hilton Barber"\end{tabular} \\ \bottomrule
\end{tabular}
\caption{Classification of errors causing missing connections.}
\label{tab:errors}
\end{table}
\section{Evaluation}
\label{sec:evaluation}
In this section, we evaluate the automatic pipeline by comparing it with the manually curated network constructed by a domain expert from a subset of the data. Note that the two networks are inherently different: the manual approach to building the social network focuses on connecting senders and receivers, while the automatic approach connects any two people mentioned together in one document and then prunes the edges with the lowest frequency scores. This method was chosen for two reasons: improving recall and potentially discovering hidden connections. Improving recall is necessary since the accuracy of automatically detecting people in letters is not 100\%, and sometimes the sender or the receiver could be overlooked: for example, some letters are only signed with a handwritten signature, and others contain OCR noise. Connecting all entities found in the documents allows to partly compensate for this, as the misssing connection can appear again in the corpus. In the case of the manual approach, the sender and receiver in noisy documents are easily derived from the context. 

With evaluating our pipeline against a manually curated network, we aim to answer two main questions:
\begin{enumerate}
    \item Which connections are present in the manual network analysis but absent from the automatically constructed network, and why?
    \item Out of the extra connections discovered by our pipeline that are not found in the manually curated sender-receiver network, how many are meaningful / useful for art history research? 
\end{enumerate}
\subsection{Missing connections}
To find and explain the connections present in the manually curated network but overlooked by our pipeline, we conduct manual evaluation on a sample of data. The sample includes 88 connections, i.e. pairs of entities, randomly selected from the manually constructed network. For each of the 176 entities, we examine every step of the pipeline – from the original images of the scanned pages to the de-duplicated set of entities found in the document – in order to identify errors. Table~\ref{tab:errors} contains the classification of errors identified during manual evaluation, and Figure~\ref{fig:pipeline_sankey} shows a Sankey chart of these errors occuring on different steps of the pipeline.  
\begin{table}[]
\begin{tabular}{@{}lll@{}}
\toprule
\textbf{case} & \textbf{count} & \multicolumn{1}{c}{\textbf{example}} \\ \midrule
\multicolumn{3}{c}{\textbf{Connections}} \\ \midrule
Direct connection found & 18 & Mieke Groot and Richard Meitner opened a studio together \\ \midrule
Possible indirect connection & 1 & \begin{tabular}[c]{@{}l@{}}Albert Lewis and Harvey Littleton: Valkema mentioned\\ Littleton in a letter to Lewis about an exhibition\end{tabular} \\ \midrule
No evidence of connection & 5 & \begin{tabular}[c]{@{}l@{}}Harvey Littleton and Hasse Holmerud are mentioned in\\ the same letter but there is no known connections between them\end{tabular} \\ \midrule
Error & 1 & \begin{tabular}[c]{@{}l@{}}Incorrect connection between Fritz Gramberg and Mart Stam:\\ the entity mention present in the document is "F. Kramer", \\ which got linked to Fritz Gramberg during record linkage\end{tabular} \\ \midrule
\multicolumn{3}{c}{\textbf{Wikidata links}} \\ \midrule
Correct & 33 & \begin{tabular}[c]{@{}l@{}}Richard Price (Dutch sculptor) was correctly linked to Q65178606,\\ despite the entity being highly ambiguous\end{tabular} \\
Wrong & 4 & \begin{tabular}[c]{@{}l@{}}Albert Lewis (American glass artist and journalist, no Wikidata page) \\ is incorrectly linked to Q119102235 (English footballer)\end{tabular}

\end{tabular}

\caption{Manual evaluation results for the extra connections detected by the pipeline.}
\label{tab:extra-connections}
\end{table}
\subsection{Extra connections}
To determine whether the extra connections discovered by our pipeline are valuable for art history research, we randomly sample 25 connections from the automatically constructed network that are not present in the manually curated network. These 25 connections (50 people in total, 37 unique) are then carefully evaluated by the domain expert, who reviews the context and annotates the interactions as meaningful or not. The results are shown in Table~\ref{tab:extra-connections}. Additionally, the domain expert reviews the Wikidata links for every person present in the sample, evaluating entity linking – the final step of our pipeline. The evaluation results are shown in Table~\ref{tab:extra-connections}.
\begin{table}[]
\centering
\begin{tabular}{@{}rrrr@{}}
\toprule
\textbf{} & \textbf{G\_automatic} & \textbf{G\_manual} & \textbf{G\_small} \\ \cmidrule(l){2-4} 
\multicolumn{1}{r|}{number of components} & 1 & 3 & 1 \\
\multicolumn{1}{r|}{number of nodes} & 868 & 124 & 43 \\
\multicolumn{1}{r|}{average node degree} & 86.54 & 2.35 & 3.86 \\
\multicolumn{1}{r|}{average weighted degree} & 306.20 & 12.74 & 11.67 \\
\multicolumn{1}{r|}{density} & 0.10 & 0.02 & 0.09 \\
\multicolumn{1}{r|}{diameter} & 5 & 4 & 4 \\
\multicolumn{1}{r|}{average clustering} & 0.85 & 0.17 & 0.49 \\
\multicolumn{1}{r|}{modularity} & 0.42 & 0.17 & 0.36 \\ \cmidrule(r){1-1}
\end{tabular}
\caption{Graph statistics of the three networks}
\label{tab:graph-stats}
\end{table}
\begin{figure}[ht]
    \centering
    \begin{subfigure}[t]{0.32\textwidth}
        \centering
        \includegraphics[width=\linewidth]{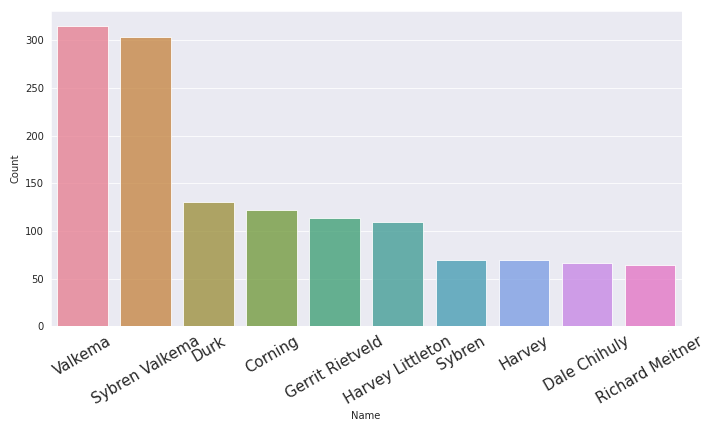}
        \caption{Top-10 entities after NER.}
        \label{fig:figure1}
    \end{subfigure}
    \hfill
    \begin{subfigure}[t]{0.32\textwidth}
        \centering
        \includegraphics[width=\linewidth]{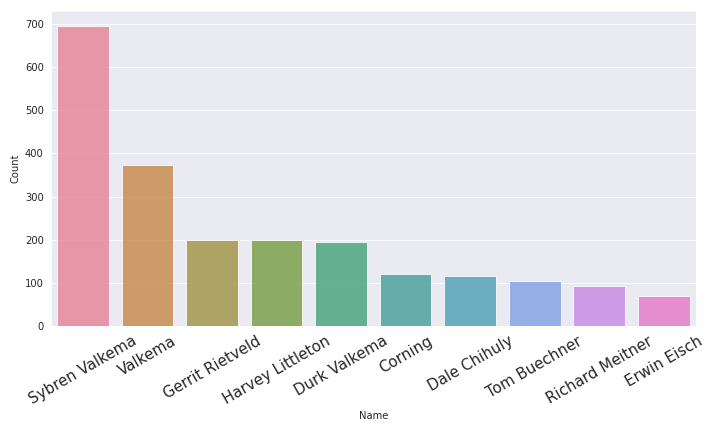}
        \caption{Top-10 entities after record linkage.}
        \label{fig:figure2}
    \end{subfigure}
    \hfill
    \begin{subfigure}[t]{0.32\textwidth}
        \centering
    \includegraphics[width=\linewidth]{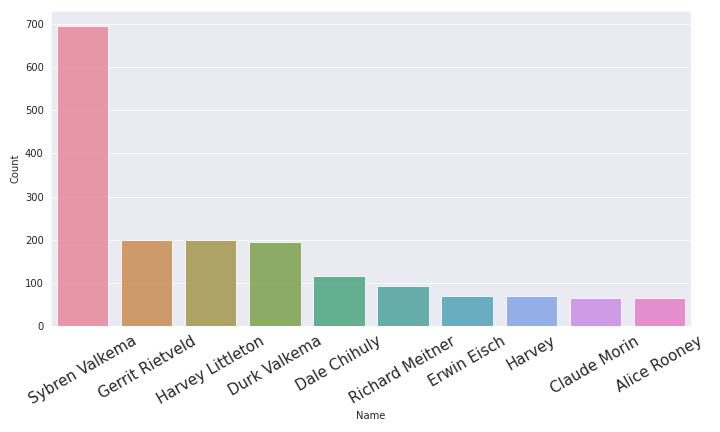}
        \caption{Top-10 entities after entity linking.}
        \label{fig:figure3}
    \end{subfigure}
    
    \caption{Top-10 most frequent unique entities in the corpus after 3 steps of the pipeline: NER, record linkage and entity linking.}
    \label{fig:top-10-pipeline}
\end{figure}
\section{Results and Discussion}
\label{sec:results-discussion}
In this section, we first discuss the strengths and limitations of the automatic network construction pipeline presented in our study. Next, we present the analysis of the three social networks constructed from the correspondence archive, comparing their structural differences, community dynamics, and centrality profiles of the most prominent nodes.
\subsection{Pipeline evaluation}
The evaluation results presented on Table~\ref{tab:extra-connections} and Figure~\ref{fig:pipeline_sankey} suggest overall strong performance of our pipeline despite certain limitations. As Figure~\ref{fig:pipeline_sankey} shows, 56 out of the 176 sampled entities were not possible to detect automatically in text and are therefore excluded from the evaluation. Out of the remaining 120 entities, 71 were processed correctly, which makes the overall accuracy 59.2\%. The biggest challenge faced by the pipeline is text recognition: 37 out of 120 entities were unrecognisable to the human annotator due to the noise introduced by OCR and HTR, making the relative accuracy of the text recognition step 69.2\%. Next step, entity recognition, performs considerably well with the relative accuracy of 94\%. Finally, the relative accuracy of record linkage on this sample is as high as 98.6\%. The evaluation was performed manually in a cascade manner due to the lack of ground truth, which makes it impossible to evaluate every step separately on a large set of data. While we recognise this limitation, we argue that the manually annotated sample is large enough to draw meaningful conclusions about the performance of the pipeline. 
Figure~\ref{fig:top-10-pipeline} shows top-10 most frequent entities after each of the 3 steps in the pipeline: NER, record linkage, and entity linking. The figure illustrates the process of gradually refining the set of entities with each step, until only the entities with the highest confidence are left for constructing social networks.

The last step of the pipeline, entity linking, is evaluated by domain expert on another sample, as seen in Table~\ref{tab:extra-connections}. Here, 33 out of 37 unique entities were correctly linked to the corresponding Wikidata pages, making the accuracy of this step 89.2\%. Additionally, an extra step of evaluating record linkage is performed on this data sample to double-check the previous evaluation results: 1 out of 37 names was linked incorrectly, making the accuracy 97.3\%, which is consistent with the number derived from Figure~\ref{fig:pipeline_sankey}. Finally, the pipeline shows strong results in discovering hidden connections: out of the 25 sampled connections, 18 are labelled as meaningful and 1 as potentially meaningful, leading to the accuracy score between 72\% and 76\%. While further research is needed to evaluate the ability of the pipeline to discover meaningful connections on larger samples of data, our evaluation results are promising. 

\subsection{Network analysis}
Table~\ref{tab:graph-stats} shows graph-based statistics of the three social networks constructed in our study. Interestingly, $G_{manual}$ contains multiple components, which means that not all letters in the archive were addressed to, or written by, Sybren Valkema. Another observation is that $G_{small}$, which was constructed from the same data subset as $G_{manual}$, contains almost 3 times as few nodes (since the names not linkable to Wikidata were filtered out) but displays considerably higher density and modularity than $G_{manual}$. In terms of density and modularity, $G_{small}$ is closer to $G_{auto}$: despite a significant difference in size, the two graphs share similar structure.

In order to gain more insights about the three networks, we turn to the centrality profiles of their top-10 most prominent nodes, as shown in Tables~\ref{tab:centrality-auto},~\ref{tab:centrality-manual} and ~\ref{tab:centrality-small}. As expected, in all three cases Sybren Valkema is by far the most important actor in the network. Durk Valkema is another individual present in top-10 for all three networks, and in $G_{small}$ his high betweenness centrality score highlights his role as a connector. In case of other actors, there is a noticeable difference between the $G_{manual}$ and the automatically constructed networks. 

Firstly, while Veronique Valkema is a prominent actor in $G_{manual}$, she does not have a Wikidata page and is therefore absent from $G_{small}$ as well as $G_{auto}$. Secondly, Gerrit Rietveld is seen in top-10 for the automatically constructed networks but absent from $G_{manual}$ due to a false positive error in NER: the entity mentioned very frequently in the corpus is "Gerrit Rietveld Academy". Lastly, Albert Lewis is present in all networks and appears in top-10 for $G_{manual}$ and $G_{small}$ but gets overshadowed by more impactful nodes in  $G_{auto}$. A likely explanation is the different nature of the data: while $G_{small}$ and $G_{manual}$ are constructed from a data subset that strictly contains letters, $G_{auto}$ was constructed from a broad set of documents found in correspondence, which includes articles and brochures. We assume that $G_{auto}$ contains more indirect professional relationships (eg., two artists mentioned together in the same article), while $G_{small}$ and $G_{manual}$ focus more on informal personal relationships (eg., one artist being mentioned to another in a conversation by mail). 

\begin{figure}
    \centering
    \begin{subtable}[t]{1\textwidth}
        \centering
        \begin{tabular}{@{}lrrrrrr@{}}
\toprule
\multicolumn{1}{l}{\multirow{2}{*}{\textbf{name}}} & \multicolumn{6}{c}{\textbf{centrality}} \\ \cmidrule(l){2-7} 
\multicolumn{1}{l}{} & \textbf{degree} & \textbf{weighted degree} & \textbf{betweenness} & \textbf{closeness} & \textbf{eigenvector} & \textbf{composite} \\ \midrule
Sybren Valkema & \textbf{1.000} & \textbf{1.000} & \textbf{1.000} & \textbf{1.000} & \textbf{1.000} & 5.000 \\
Harvey Littleton & {\ul 0.679} & \textbf{0.569} & 0.154 & {\ul 0.722} & {\ul 0.747} & 2.871 \\
Gerrit Rietveld & 0.607 & 0.397 & {\ul 0.174} & 0.672 & 0.570 & 2.420 \\
Dale Chihuly & 0.615 & 0.550 & 0.061 & 0.668 & 0.438 & 2.331 \\
Erwin Eisch & 0.621 & 0.341 & 0.095 & 0.680 & 0.463 & 2.201 \\
Durk Valkema & 0.507 & 0.364 & 0.133 & 0.613 & 0.525 & 2.142 \\
Claude Morin & 0.477 & 0.368 & 0.051 & 0.595 & 0.542 & 2.033 \\
Marvin Lipofsky & 0.601 & 0.229 & 0.074 & 0.668 & 0.269 & 1.841 \\
Joel Myers & 0.462 & 0.302 & 0.040 & 0.586 & 0.363 & 1.754 \\
Lino Tagliapietra & 0.510 & 0.314 & 0.028 & 0.603 & 0.222 & 1.678 \\ \bottomrule
\end{tabular}
        \caption{Top-10 nodes in $G_{auto}$.}
        \label{tab:centrality-auto}
    \end{subtable}
    
    \vspace{0.5cm} 
    
    \begin{subtable}[t]{1\textwidth}
        \centering
        \begin{tabular}{@{}lrrrrrr@{}}
            \toprule
            \multicolumn{1}{l}{\multirow{2}{*}{\textbf{name}}} & \multicolumn{6}{c}{\textbf{centrality}} \\ \cmidrule(l){2-7} 
            \multicolumn{1}{l}{} & \textbf{degree} & \textbf{weighted degree} & \textbf{betweenness} & \textbf{closeness} & \textbf{eigenvector} & \textbf{composite} \\ \midrule
            Sybren Valkema & \textbf{1.000} & \textbf{1.000} & \textbf{1.000} & \textbf{1.000} & \textbf{1.000} & 5.000 \\
            Harvey Littleton & {\ul 0.393} & \textbf{0.360} & 0.126 & {\ul 0.550} & {\ul 0.626} & 2.055 \\
            Albert Lewis & 0.321 & 0.324 & 0.031 & 0.431 & 0.503 & 1.611 \\
            Durk Valkema & 0.286 & 0.180 & \textbf{0.255} & 0.458 & 0.154 & 1.334 \\
            Gerrit Rietveld & 0.143 & 0.180 & 0.121 & 0.367 & 0.487 & 1.299 \\
            Erwin Eisch & 0.214 & 0.225 & 0.013 & 0.392 & 0.422 & 1.267 \\
            Sam Herman & 0.286 & 0.189 & 0.085 & 0.417 & 0.283 & 1.260 \\
            Finn Lynggaard & 0.143 & 0.090 & 0.007 & 0.344 & 0.193 & 0.776 \\
            Claude Morin & 0.143 & 0.081 & 0.002 & 0.367 & 0.177 & 0.770 \\
            Mieke Groot & 0.143 & 0.090 & 0.038 & 0.392 & 0.089 & 0.752 \\ \bottomrule
        \end{tabular}
        \caption{Top-10 nodes in $G_{small}$.}
        \label{tab:centrality-small}
    \end{subtable}
    
    \vspace{0.5cm} 
    
    \begin{subtable}[t]{1\textwidth}
        \centering
         \begin{tabular}{@{}lrrrrrr@{}}
\toprule
\multirow{2}{*}{\textbf{name}} & \multicolumn{6}{c}{\textbf{centrality}} \\ \cmidrule(l){2-7} 
 & \textbf{degree} & \textbf{weighted degree} & \textbf{betweenness} & \textbf{closeness} & \textbf{eigenvector} & \textbf{composite} \\ \midrule
Sybren Valkema & \textbf{1.000} & \textbf{1.000} & \textbf{1.000} & \textbf{1.000} & \textbf{1.000} & 5.000 \\
Veronique Valkema & {\ul 0.161} & \textbf{0.173} & 0.012 & {\ul 0.548} & {\ul 0.377} & 1.271 \\
Durk Valkema & 0.025 & 0.074 &  0.000 & 0.514 & 0.381 & 0.996 \\
Jaroslava Brychtova & 0.017 & 0.046 & 0.008 & 0.512 & 0.282 & 0.865 \\
Stanislav Libensky & 0.017 & 0.044 & 0.008 & 0.512 & 0.270 & 0.852 \\
Charles Bray & 0.008 & 0.032 & 0.000 & 0.508 & 0.247 & 0.796 \\
Ingeborg Lundin & 0.008 & 0.032 & 0.000 & 0.508 & 0.211 & 0.759 \\
Albert Lewis & 0.000 & 0.026 & 0.000 & 0.506 & 0.222 & 0.754 \\
Kl.Laansma & 0.008 & 0.026 & {\ul 0.017} & 0.510 & 0.188 & 0.748 \\
Arjan Muijlwijk & 0.008 & 0.032 & 0.000 & 0.508 & 0.196 & 0.745 \\ \bottomrule
\end{tabular}
        \caption{Top-10 nodes in $G_{manual}$}
        \label{tab:centrality-manual}
    \end{subtable}
    
    \caption{Centrality profiles of the top-10 most prominent nodes in the three networks.}
    \label{fig:centrality-profiles}
\end{figure}
\section{Conclusion and Future work}
\label{sec:conclusion-future}
We presented a pipeline approach for constructing social networks from heterogeneous archival data, using the correspondence archive of Sybren Valkema as a case study. Our method is evaluated against a sender-receiver network manually curated by a domain expert. The evaluation results are promising, with text recognition on scanned pages identified as the main challenge. The pipeline has shown strong performance in identifying the sender-receiver connections across the corpus of correspondence, as well as discovering additional connections deemed meaningful from an art history perspective. Additionally, our contribution highlights the importance of interdisciplinary collaborations in digital humanities: the research was conducted by a team comprising an expert in art history (Carlotta Capurro) and two experts in natural language processing (Vera Provatorova and Evangelos Kanoulas).
\subsection{Future work}
We envision the following future work directions for our contribution:
\begin{itemize}
    \item Extending entity linking to domain-specific knowledge bases. Some entities in the correspondence archive of Sybren Valkema are absent from Wikidata but present in the Art\&Architecture Thesaurus\footnote{\url{https://www.getty.edu/research/tools/vocabularies/aat/index.html}} or the RKDartists database\footnote{\url{https://data.netwerkdigitaalerfgoed.nl/rkd/rkdartists}}. Currently these entities are excluded from the final network, so incorporating additional information sources could help in discovering new connections.
    \item Using entity linking to add extra features to the nodes. In the current analysis, we used country information derived from Wikidata to highlight interactions between Dutch and American artists. Adding more features could further enhance the analysis: for instance, classifying nodes by profession (artist, art collector, journalist, etc.) could provide additional insights.
    \item Improving text recognition, especially for handwritten pages: as this step has shown to be the most challenging for our pipeline, we expect that better HTR and OCR results will improve the resulting network analysis. One way to improve HTR would be to fine-tune the text recognition model on the handwritings present in the archive, since prior work shows that handwriting style is an important feature for HTR~\cite{capurro2023experimenting}.
    \item Adding a temporal dimension to the analysis: given that the correspondence archive of Sybren Valkema is spanning several decades, studying the evolution of the social network over time could provide valuable insights about its dynamics.
\end{itemize}
\bibliography{anthology,references}
\bibliographystyle{splncs04}




\end{document}